# Increasing altruistic and cooperative behaviour with simple moral nudges

Valerio Capraro[1], Glorianna Jagfeld[2], Rana Klein[3], Mathijs Mul[3] & Iris van de Pol[3]

[1]Middlesex University of London. [2]Universität Stuttgart. [3]Universiteit van Amsterdam.

Contact Author: V.Capraro@mdx.ac.uk

**Abstract**

The conflict between pro-self and pro-social behaviour is at the core of many key problems of our time, as, for example, the reduction of air pollution and the redistribution of scarce resources. For the well-being of our societies, it is thus crucial to find mechanisms to promote pro-social choices over egoistic ones. Particularly important, because cheap and easy to implement, are those mechanisms that can change people's behaviour without forbidding any options or significantly changing their economic incentives, the so-called "nudges". Previous research has found that moral nudges (e.g., making norms salient) can promote pro-social behaviour. However, little is known about whether their effect persists over time and spills across context. This question is key in light of research showing that pro-social actions are often followed by selfish actions, thus suggesting that some moral manipulations may backfire. Here we present a class of simple moral nudges that have a great positive impact on pro-sociality. In Studies 1-4 (total N = 1,400), we use economic games to demonstrate that asking subjects to self-report "what they think is the morally right thing to do" does not only increase pro-sociality in the choice immediately after, but also in subsequent choices, and even when the social context changes. In Study 5, we explore whether moral nudges promote charity donations to humanitarian organisations in a large (N=1,800) crowdfunding campaign. We find that, in this context, moral nudges increase donations by about 44 percent.

**Introduction**

Many decision problems involve a conflict between what is good for oneself and what is good for others. For example, reducing $CO_2$ emissions, recycling, and donating to charity are all *pro-social* choices that are individually costly: they require people to pay a personal cost to benefit other people or the society as a whole[1-4]. Since individuals have an incentive to maximise their own benefit, finding mechanisms to reduce free-riding and promote pro-social behaviour is of great importance for the well-being of our society[5-12].

One strand of literature has focused on mechanisms such as punishing free-riders[13-16], rewarding pro-social actors[17-19], and the interplay between these two[20-23]. While these approaches have been shown to promote pro-sociality, and punishment has been adopted by most countries to sanction free-riders, their drawback is that they are costly. For example, in order to punish self-regarding behaviour, institutions must first pay a cost to monitor people's behaviour and find out who acted selfishly and who did not. Then they must pay a cost to punish those who acted selfishly. A similar drawback holds for rewarding pro-social actors. In order to reward pro-social actions, institutions must first pay a cost to monitor people's actions and find out who acted pro-socially and who did not; then they must pay a cost to reward those who acted pro-socially.

Thus, particularly important, from the point of view of creating societal benefit, are those mechanisms that can change people's behaviour without forbidding any options or significantly changing their economic incentives – the so-called "nudges".[24-25] They are cheap and easy to implement, because they allow to avoid (i) the direct cost of changing people's economic incentives and/or limiting people's action space, (ii) the monitoring cost of finding out which choice each individual made and, possibly, the cost of punishing or rewarding each choice, and (iii) the technical difficulties associated with finding out individual choices. Thus, finding simple nudges that are able to increase pro-social behaviour is of great importance and could have impactful implications on policy design.

A large amount of research testifies of the considerable impact that norms can have on human behaviour in situations as diverse as charity donations[26], alcohol consumption[27], littering[28], water consumption[29], risky driving[30], and many others. In short, this research suggests that people tend to follow the choice that they perceive to be the norm in a given context, which, in turn, suggests that simple reminders that make the morality of an action salient may have a positive impact on pro-social behaviour. Along these lines, for example, it has been shown that religious reminders favour honesty[31] and cooperation[32-33] in economic games. Furthermore, changing the names of the available actions to make them seem less or more moral has a large effect on people's behaviour in decision problems where there is a tension between equity and efficiency[34-35].

Inspired by this line of research, in this paper, we explore the effect of simple "moral nudges" on pro-social behaviour in the context of economic games. In our typical experiment, participants have to decide between a pro-self and a pro-social course of action, but, before making their actual choice, they are asked what they think is the morally right thing to do. Besides investigating whether answering this question makes participants more pro-social in the given interaction, we also examine whether its effect persists over time and spills across contexts.

While previous research has consistently found that moral nudges and, more generally, norm-based policy interventions persist at least for some time[33,36-37], there is a lot of uncertainty concerning their spillover effect across contexts. More specifically, to the best of our knowledge, no study has explored spillover effects in the domain of simple moral nudges as the ones we explore in this work. In the more general domain of policy interventions, previous empirical studies have led to mixed results, as they offered evidence of both negative and positive spillovers. In the realm of negative spillovers, previous research has shown that people donate less to charity directly after making a choice they consider moral than after a choice they consider immoral[38], that "recalling one's own (im)moral behaviour leads to compensatory rather than consistent moral action"[39], that "people act less altruistically and are more likely to cheat and steal after purchasing green products than after purchasing conventional products"[40], that "people are more willing to express attitudes that could be viewed as prejudiced when their past behaviour has established their credentials as nonprejudiced persons"[41]. See also refs 42-45. In the realm of positive spillovers, it has been shown that costly investments in social image generate positive spillovers in terms of generosity later on[46], that previously established cooperative precedents lead to higher cooperation in subsequent games[47-50], that cooperative behaviour learned in a particularly cooperative setting spills over to a number of other settings involving pro-sociality[50], and that different pro-social behaviours are all positively correlated[51-54].

In this work, we first demonstrate that results from previous related literature extend to the domain of our moral nudges. In particular, we show that moral nudges have a positive effect on the choice immediately after the nudge (Study 1-2) and that their effect persists to a second interaction (Study 3). Then, we go beyond previous literature by showing that the effect of moral nudges spill across contexts (Study 4).

Additionally, we also explore whether our moral nudges can be used to increase charity donations to real humanitarian organisations. The motivation for exploring this question comes from the fact that whether behaviour in economic games reflects behaviour in real life or not is still debated, with some studies finding a positive association[55-59], whereas others challenge this view by reporting only a weak association, if any at all[60-62]. Thus, to strengthen our conclusions, we also provide a demonstration that moral nudges can be used to increase charity donations to real humanitarian organisations: in Study 5, we find that, in our setting, moral nudges significantly increase charity donations by 44%.

**Study 1**

We begin by exploring whether moral nudges are effective in increasing altruistic behaviour in the Dictator Game (DG), a simple economic game standardly used to measure individuals' altruistic attitudes[59,63].

For this and the following experiments, informed consent is obtained by all participants and their anonymity is ensured. The experiments were carried out in accordance with relevant guidelines and regulations. These experiments were conducted when all authors were affiliated to the University of Amsterdam. According to the Dutch legislation, this is a non-NWO study, that is (i) it does not involve medical research and (ii) participants are not asked to follow rules of behaviour. See http://www.ccmo.nl/ attachments/files/wmo-engelse-vertaling-29-7-2013-afkomstig-van- vws.pdf, §1, Article 1b, for an English translation of the Medical Research Act. Thus (see http://www.ccmo.nl/en/non-wmo- research) the only legislations which apply are the Agreement on Medical Treatment Act, from the Dutch Civil Code (Book 7, title 7, §5), and the Personal Data Protection Act (a link to which can be found in the previous webpage). All the studies presented in this work conform to both regulations.

*Subjects*

We recruit N = 300 subjects living in the US using the online labour market Amazon Mechanical Turk (AMT). AMT experiments are easy and cheap to implement, because subjects participate from their homes by simply completing an online incentivized survey that takes no more than a few minutes. This allows researchers to significantly decrease the stakes of the experiment, without compromising the results. Several studies have indeed shown that data gathered using AMT are of no less quality than data gathered using the standard physical lab[64-69]. Moreover, as an upside with respect to standard laboratory experiments, AMT experiments use samples that are more heterogeneous than the standard laboratory experiments, that are typically conducted using a pool of students[67]. Other studies have pointed out to potential problems in collecting data on AMT, including the presence of experienced subjects, who may be unaffected by experimental manipulations and thus may decrease effect sizes[70] or the presence of AMT workers using Virtual Private Servers (VPS) to participate multiple times in an experiment[71]. To increase data quality, we recruited subjects with an AMT approval rate greater than 90% (AMT keeps track of this information and allows experimenters to filter subjects accordingly) and we also asked subjects comprehension questions during the survey to make sure that they understood the crucial parts of the experiment. In particular, we asked which choice maximizes the subject's payoff and which choice maximizes other subjects' payoff. In this way, we made sure that subjects who pass the comprehension questions have a clear understanding of the conflict between the pro-self and the pro-social choice that characterizes our decision problems. Subjects who failed the comprehension question were automatically excluded from the survey. Comprehension questions were asked before the moral nudge manipulation as to avoid any risk of differential attrition across conditions. In order to minimize the effect of potential duplicates, after collecting the data, we checked for multiple IP addresses and multiple AMT IDs. In those cases where we found any duplicates, we kept only the first observation (as determined by the starting date) and we discarded the remaining ones. We did not take any measure of experience in AMT experiments to identify a possible decrease in effect sizes due to the presence of experienced subjects. Thus, our results are likely to represent a lower bound of the true effect sizes.

*Procedure*

After entering their AMT IDs, subjects are given 20c and are told they are paired with another subject who is given nothing. They are also told that they can donate any part of their endowment to the other subject (available choices: 0c, 2c, 4c, …, 18c, 20c). To make sure they understand the situation, we ask all subjects two comprehension questions regarding what choice would maximise their payoff and what choice would maximise the other person's payoff. Only those who answer both comprehension questions correctly are allowed to proceed to the experiment. Subjects are randomly divided in three conditions: *DG*, *DGpersonal*, *DGdescriptive*. The *DG* condition is the baseline: subjects are asked how much, if any, they want to donate to the other person. In the *DGpersonal* condition, right before asking subjects to state their donation, we ask them: "what do you personally think is the morally right thing to do in this situation?" In the *DGdescriptive* condition, right before asking subjects to state their donation, we ask them: "what do you think your society considers to be the morally right thing to do in this situation?" After making a choice, subjects enter a standard demographic questionnaire. After the questionnaire, subjects receive the completion code of the survey through which they can claim their bonus, that is paid on top of the participation fee (that is 40c). We refer to the Supplementary Material, Part SM2, for full experimental instructions.

*Results*

Excluding subjects who did not pass the comprehension questions or completed the survey more than once (for whom we keep only the first observation), we remain with N=282 subjects. Exclusion rates in this study and the following studies are in line with those of other published studies using the DG and the PD[65]. The results are visualised in Figure 1, where "% pro-sociality" stands for average percentage of the endowment donated by all players in the given condition. Since linear regression predicting "pro-sociality" as a function of a dummy variable that represents whether a subject participated in the *DGpersonal* condition or in the *DGdescriptive* condition finds no significant difference between *DGpersonal* and *DGdescriptive* (p = 0.473), we merge these two conditions to form the *DGnudged* condition. When nudging subjects, average percentage of the endowment donated increases from 21.2% in the DG to 30.6% in the *DGnudged*. The increase is statistically significant, as shown by linear regression predicting pro-sociality as a function of a dummy variable that represents whether an individual participated in *DGnudged* or in the standard *DG* (p = 0.005). Thus, our first study shows that moral nudges have a positive impact on altruistic behaviour in the standard DG (see Supplementary Material, Table SM1, for regression details).

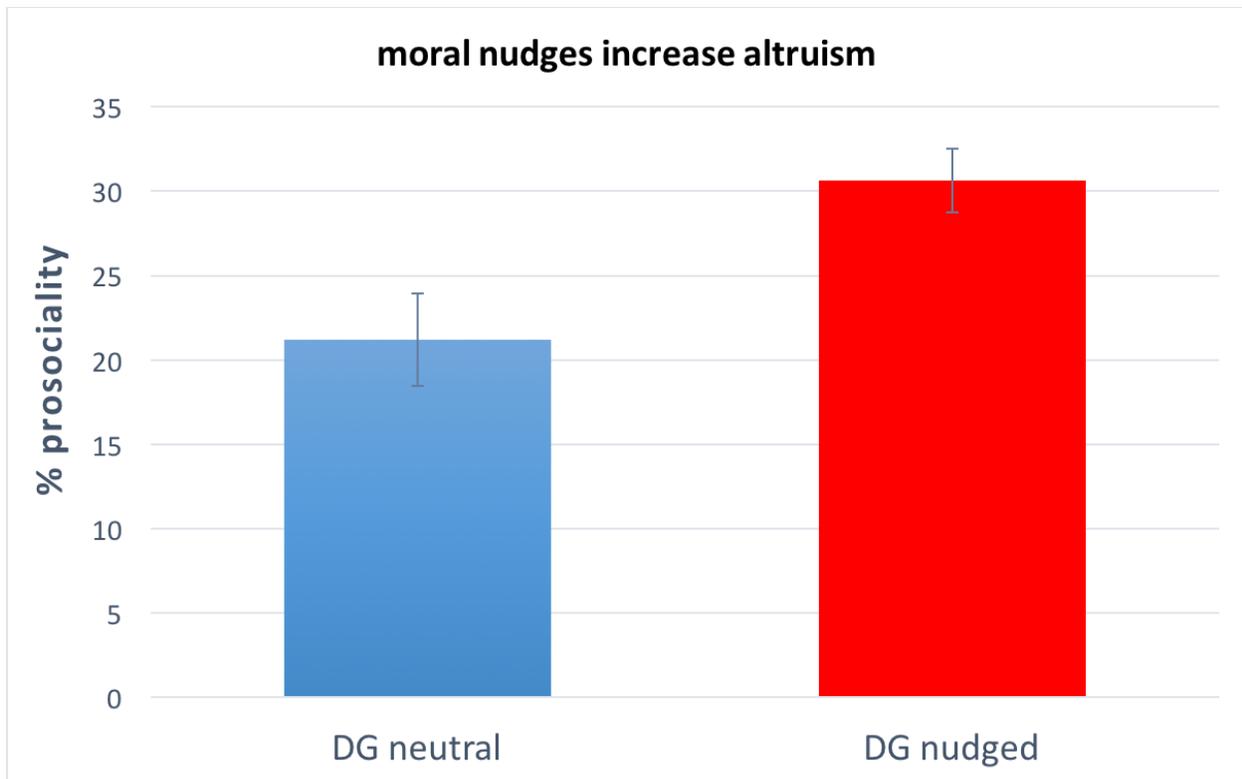

**Figure 1: Moral nudges increase altruistic behaviour in Dictator games.** *Results of Study 1 (N=282). Subjects who are asked what they personally think is the morally right thing to do or what they think their society considers to be the morally right thing to do before playing a DG (i.e., subjects in the "DG nudged" condition), donate significantly more than subjects in the baseline DG (p=0.005). The y-axis represents the average percentage of the endowment donated by subjects in the corresponding condition. Error bars represent +/- Standard Error of the Mean (SEM). We refer to the Supplementary Material, Table SM1, for regression details.*

**Study 2**

Next, we examine whether moral nudges have a positive effect also on cooperative behaviour. We design a new experiment in which the DG is replaced by a Prisoner's Dilemma (PD), an economic game standardly used to measure individuals' cooperative attitudes[56,72].

*Subjects*

We recruit N = 300 subjects living in the US and with an AMT approval rate greater than 90% using the online labour market AMT. None of these subjects participated in the previous study.

*Procedure*

This study is identical to Study 1, apart from two changes. The first one is that the Dictator Game is replaced by a Prisoner's Dilemma (PD). In the PD, subjects are given 10c and are asked whether they want to keep it or give it to the other person. In the latter case, the 10c are doubled and earned by the other person. Importantly, participants are informed that the other person is facing exactly the same set of instructions. Thus, each participant gets a better payoff if they keep the money (defect), but, if both participants keep the money, then they only get 10c each, which is less than the 20c each that they would get if they both gave the money away (cooperate). Previous research suggests that cooperative behaviour is different from altruistic behaviour; specifically, people who act altruistically in DG typically cooperate in PD, but not the converse[51]. The second change is that, in order to participate in the experiment, subjects have to answer four comprehension questions, instead of two. We refer to the Supplementary Material, Part SM2, for full experimental instructions.

*Results*

Excluding subjects who did not pass the comprehension questions or completed the survey more than once, we remain with N=257 subjects. As before, we build *PDnudged*, by merging *PDpersonal* with *PDdescriptive*, because linear regression shows that the rate of cooperation in the *PDpersonal* condition is not statistically different from the rate of cooperation in the *PDdescriptive* condition (p = 0.327). Results are visualised in Figure 2, where "% pro-sociality" stands for the average percentage of one's own endowment given to the other player. It is clear that results are in line with those of the previous experiment: the baseline leads to the lowest average giving (32.9%), while *PDnudged* give rise to a significantly higher levels of giving (48.0%, linear regression, p=0.023). (Despite the fact that cooperation in the Prisoner's dilemma is a binary variable, we decided to use linear regression instead of logistic regression to favour comparability of effect sizes across studies (see Study 4). In any case, the results remain qualitatively the same when using logistic regression instead of linear regression. In particular, instead of getting p=0.023, we get p=0.024, when using logistic regression). Thus, our second study shows that moral nudges have a positive impact on cooperative behaviour in standard economic games (see Supplementary Material, Table SM2, for regression details).

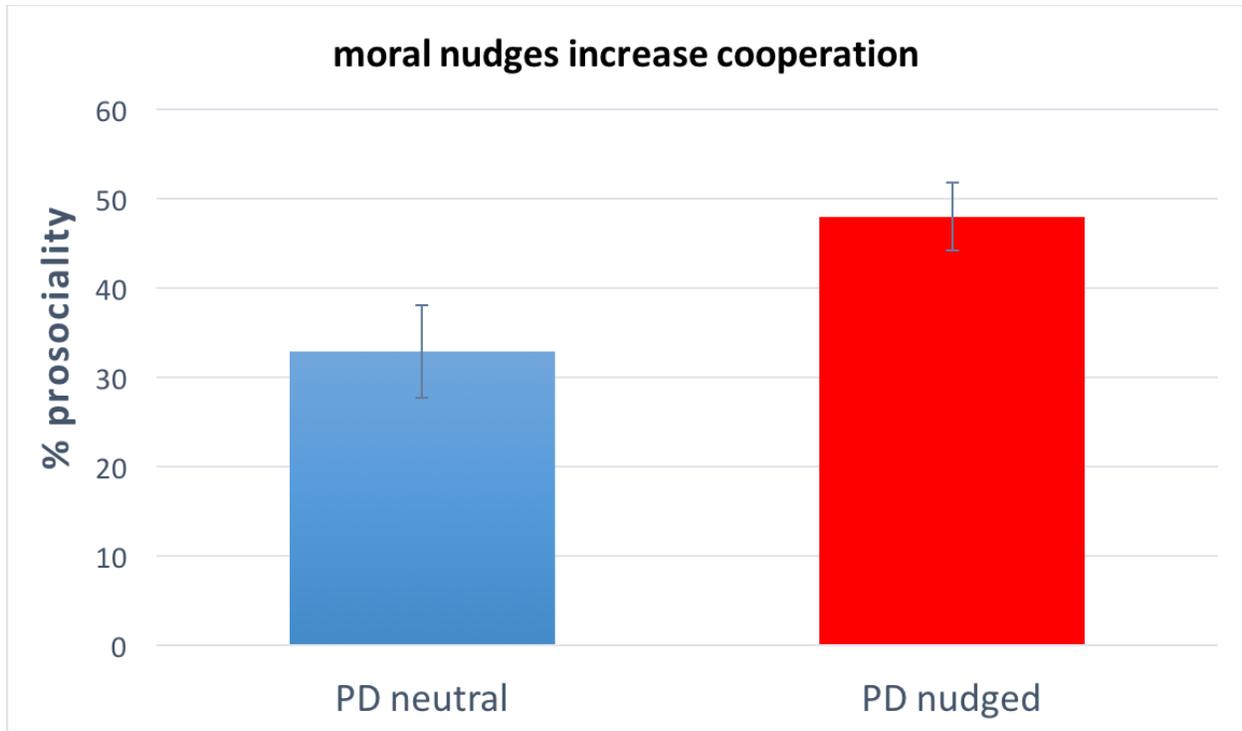

**Figure 2: Moral nudges increase cooperative behaviour in the Prisoner's Dilemma.** *Results of Study 2 (N=257). Subjects who are asked what they personally think is the morally right thing to do or what they think their society considers to be the morally right thing to do before playing a PD (i.e., subjects in the "PD nudged" condition), give significantly more than subjects in the baseline PD (p=0.023). The y-axis represents the average percentage of the endowment given by subjects in the corresponding condition. Error bars represent +/- SEM. We refer to the Supplementary Material, Table SM2, for regression details.*

**Study 3**

Then we investigate whether the positive effect of moral nudges is limited to the choice immediately after the nudge, or persists to a subsequent interaction. In this study we focus only on altruistic behaviour for two reasons: first, Dal Bó & Dal Bó[33] have already shown that a similar moral nudge persists over time in a repeated PD; second, Study 1 and Study 2 suggest that our moral nudges work similarly in the PD and the DG, therefore, although we cannot logically infer that our moral nudges equally persist to a second round both in the DG and in the PD, we expect no major differences. And, moreover, we focus only on the "personal" norm since, as shown in Studies 1 and 2, the "social" norm gives rise to similar results. Thus, we expect no major differences between the two norms.

*Subjects*

We recruit N = 200 subjects living in the US and with an AMT approval rate greater than 90% using the online labour market AMT. None of these subjects participated in the previous study.

*Procedure*

Subjects played a two-stage experiment as follows. In *Stage 1*, they face a DG in either of two conditions: *DG* and *DGpersonal*, identical to those in the first study. In *Stage* 2, all subjects play a standard non-nudged dictator game, denoted *DG2*, but with slightly different instructions than the first-stage DG: dictators are given 40c (instead of 20c) and the available donations are multiple of 4c (instead of 2c). We opt for this variant of the DG in order to avoid that people anchor their second choice to the first one[73], causing a confound that would be in the same direction as the effect that we expect to find.

*Results*

Excluding subjects who did not pass the comprehension questions or completed the survey more than once, we remain with N=172 subjects. Figure 3 shows that the percentage of altruism in *DG2* after *DG* is significantly lower than the percentage of altruism in *DG2* after *DGpersonal* (18.0% vs 25.5%; linear regression predicting DG2 choice as a function of a dummy variable that represents whether in the first stage a participant participated in *DG* or *DGpersonal*, p = 0.019). Thus, our third study shows that the positive impact of moral nudges on altruistic behaviour is not limited to the choice immediately after the nudge, but persists to at least one other choice (see Supplementary Material, Table SM3, for regression details).

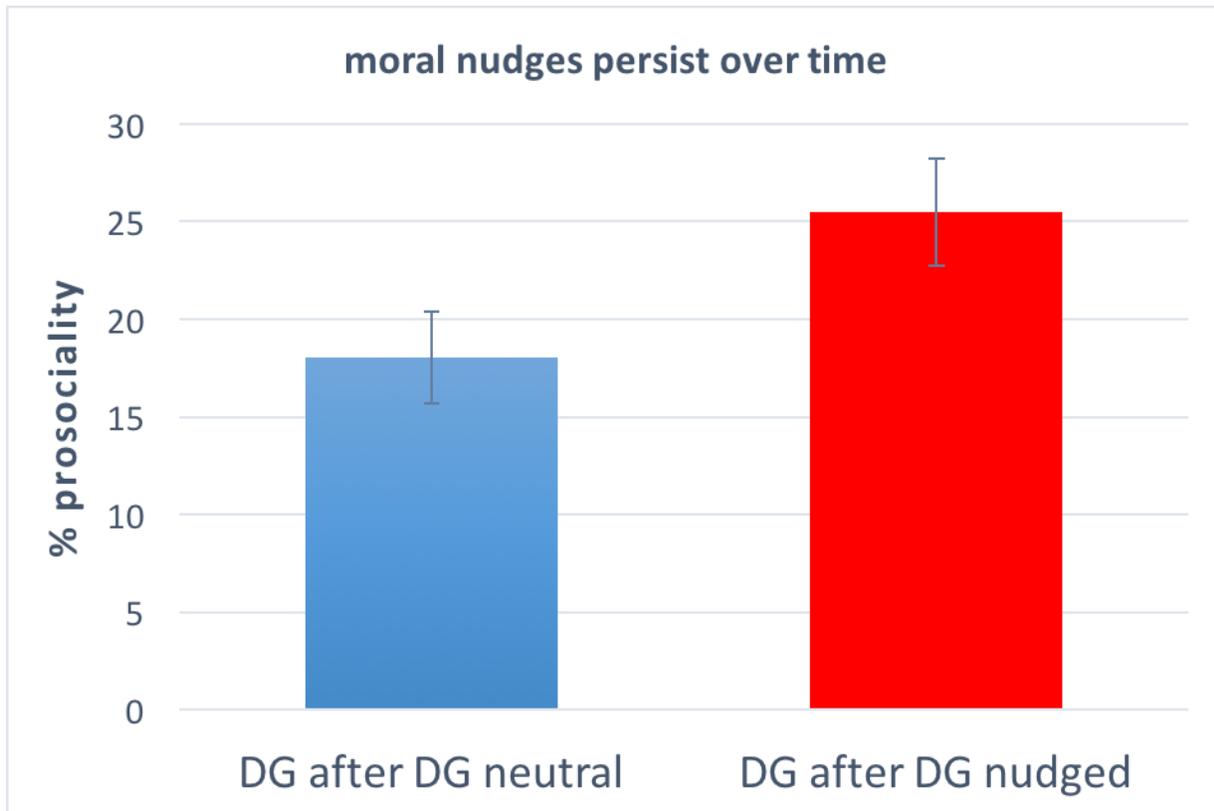

**Figure 3: Moral nudges increase altruistic behaviour in a subsequent Dictator Game.**
*Results of Study 3 (N=172). The positive effect of moral nudges on altruistic behaviour in the*

*DG is not limited to the choice immediately after the nudge, but persists to a second-stage DG, such that subjects who are nudged in Stage 1 DG donate, in a second stage DG, more than those who are not nudged (p = 0.019). The y-axis represents the average percentage of the endowment donated by subjects in the corresponding condition. Error bars represent +/- SEM. We refer to the Supplementary Material, Table SM3, for regression details.*

**Study 4**

Next, we explore whether the positive effect of moral nudges spills across contexts, in the sense that a moral nudge in the DG impacts also a subsequent PD and, conversely, a moral nudge in the PD impacts also a subsequent DG.

*Subjects*

We recruit N = 600 subjects living in the US and with an AMT approval rate greater than 90% using the online labour market AMT. None of these subjects participated in the previous study.

*Procedure*

Subjects are randomly divided in six conditions: *DG-PD*, *DGpersonal-PD*, *DGdescriptive-PD*, *PD-DG*, *PDpersonal-DG*, and *PDdescriptive-DG*, where the notation X-Y means that subjects first play game X and then play game Y. The various DG and PD conditions were identical to those in the previous studies. The DG was conducted with an endowment of 20c and the available choices were: 0c, 2c, 4c, …, 18c, 20c.

*Results*

Excluding subjects who did not pass the comprehension questions or completed the survey more than once, we remain with N=537 subjects. Direct effects of the moral nudges in Stage 1 behaviour are in line with those in the previous studies. We refer to the Supplementary Material, Table SM4, for statistical details. Here we focus on the spillover effects. Figure 4 compares the percentage of the amount given in Stage 2 after the neutral condition with the percentage of the amount given in Stage 2 after the nudged conditions (i.e., in the "neutral" column we collapse Stage 2 behaviour in PD-DG with Stage 2 behaviour in DG-PD, of course after renormalizing the data such that Stage 2 behaviour is expressed as a proportion of the amount given to the other player; similarly, in the "nudged" column, we collapse second-game behaviours in the conditions PDdescriptive-DG, PDpersonal-DG, DGdescriptive-PD, DGpersonal-PD). Figure 4 provides a first piece of evidence that moral nudges spill across contexts, as the average pro-sociality in Stage 2 increases from 30.9% to 36.8% when Stage 2 is preceded by a nudge compared to when it is not. To formally show the existence of an overall spillover effect, we use standard meta-analytic techniques as follows. First of all, we use second-stage linear regression, in which the effect of moral nudges on Stage 2 behaviour is described as a proportion of the effect of moral nudges on Stage 1, to compute the single effects. When taken singularly, the effects are not significant (*PD* after *DG* vs *PD* after *DGpersonal*: p = 0.110; *PD* after *DG* vs *PD* after *DGdescriptive*: p = 0.309; *DG* after *PD* vs *DG* after *PDdescriptive*: p = 0.348; *DG* after *PD* vs *DG* after *PDpersonal*: p = 0.889;). The fact that each of these effects is not significant is not surprising, as one may expect that the spillover effect across conditions, if present, will only constitute a proportion of the direct effect. Indeed, to shed light on whether a significant proportion of the direct effect spills over across conditions, we conduct a meta-analysis of the four effect sizes (using the Stata command: *metan effect standard_error*). In doing so, we indeed find a statistically significant overall effect, whose effect size is also relatively large (overall effect size = 0.633, 95% CI [0.047,1.219], Z = 2.12, p = 0.034). In sum, our fourth study shows that a significant proportion of the original direct effect of moral nudges on first-stage behaviour spills over across contexts in second-stage behaviour (see Supplementary Material, Table SM5, for regression details, and Supplementary Material, Figure SM1, for the forest plot of the meta-analysis).

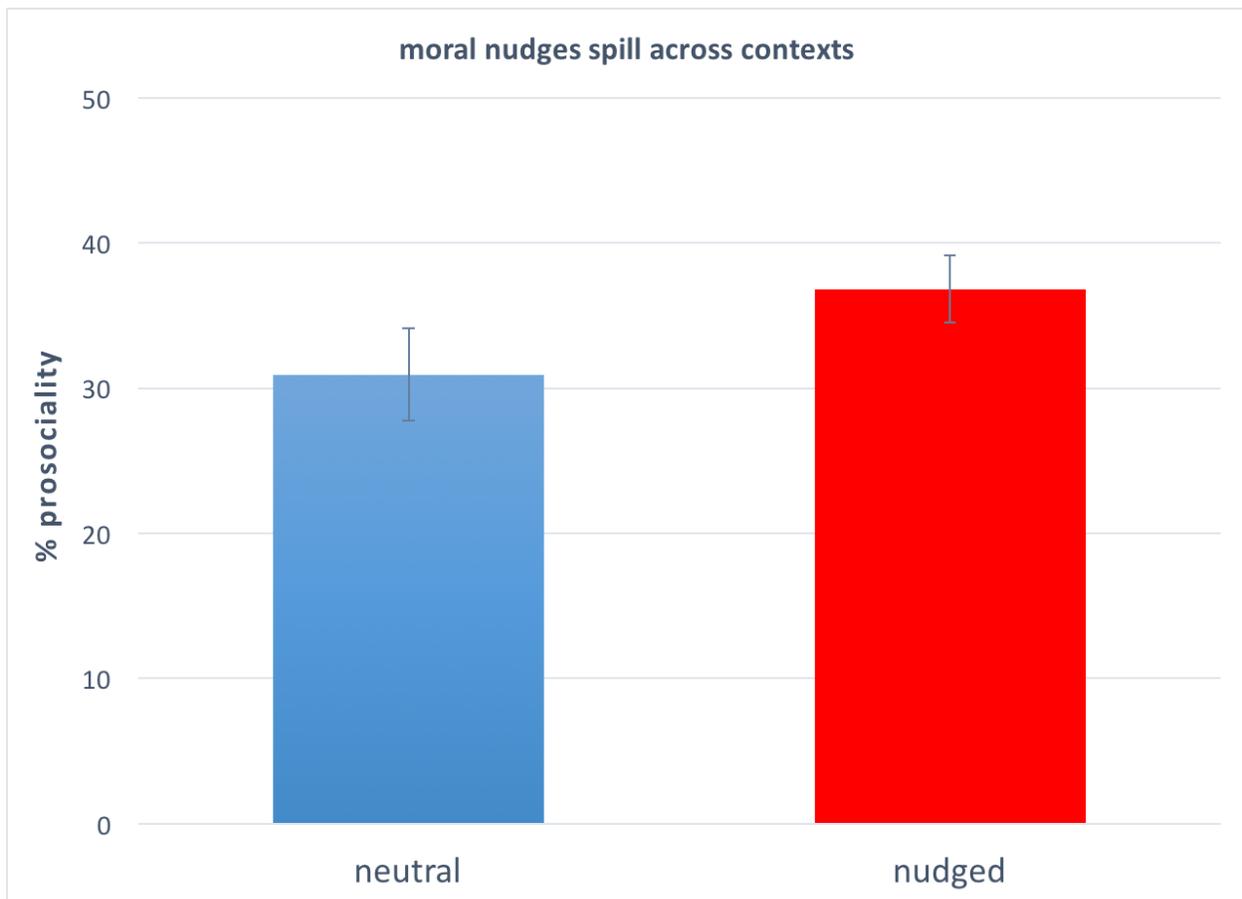

**Figure 4: Moral nudges spill across contexts.** *Results of Study 4, second stage game (N=537). Subjects who are nudged in a first-stage DG (resp. PD) tend to be more pro-social in a subsequent non-nudged PD (resp. DG) than those who are not nudged. A meta-analysis of the four conditions yields a significant positive spillover effect whose size is also relatively large (overall effect size = 0.633, 95% CI [0.047,1.219], Z = 2.12, p = 0.034), providing evidence that a significant proportion of the effect of moral nudges on first-stage behaviour spills across contexts. The y-axis represents the average percentage of the endowment given by subjects in second-stage game in the corresponding first-stage condition. Error bars represent +/- SEM. We refer to the Supplementary Material, Table SM5, for regression details, and Figure A1, for the forest plot of the meta-analysis.*

**Study 5**

We now provide a demonstration that our mechanism can be used to increase charity donations to real humanitarian organisations, at least in the context of a crowdfunding campaign conducted on Amazon Mechanical Turk.

*Subjects*

We recruit N = 1,800 subjects living in the US and with an AMT approval rate greater than 90% using the online labour market AMT. None of these subjects participated in the previous studies.

*Procedure*

We conduct three sessions, differing by small details, as detailed in the next paragraph, with the same basic structure. First, all subjects complete a 5-minute survey for 50c (the participation fee). Then they are divided between two conditions: a baseline condition and a moral nudge condition (see next paragraph for more details about the implementation of the nudges). Finally, they are divided between two more conditions, *Emergency* and *Give for France*. In the *Emergency* condition they are asked whether they want to donate part of their 50c to Emergency, a humanitarian NGO that provides emergency medical treatment to victims of war. In the Give for France condition, they are asked whether they want to donate part of their 50c to Give for France, an organization in support of the victims of the July 14, 2016 Nice attack. These experiments were conducted between July 15 and July 17, 2016.

The three sessions differ only in two aspects. In Session 1, subjects can donate only 0c or 50c (i.e., they either donate the whole participation fee, or they donate nothing) while in Sessions 2 and 3 they can donate any part of their 50c (i.e., the money they donate are still taken from the participation fee, but they can decide to donate any amount between 0c and 50c). Subjects who decide to donate a part of their 50c are asked to abandon the current AMT survey and join a "bonus survey" in which we pay the amount that they decide not to donate as a bonus. No deception is used and the donations are really sent to the corresponding organisations. The second difference regards the way the nudge is implemented. *Session 1*. In the *baseline* condition, after the 5-minute survey, subjects are told: "thanks for answering our questions"; in the *nudged* condition, subjects are asked what they think is the morally right thing to do when they see a stranger in need (available answers: help/don't help). *Session 2.* The baseline condition is the same as Session 1. In the *nudged* condition, participants are shown the instructions of a DG and are asked what they think is the morally right thing to do in that situation. Comprehension questions on the DG are *not* asked. *Session 3*. Identical to Session 2, but we add two comprehension questions for the DG (the same as in Study 1). In order to avoid selection effects, we add two comprehension questions also in the baseline condition, that are mathematically equivalent to the questions used in the nudged condition. Specifically, we ask subjects what number x would maximise the expression $20 - x$, and what number would minimise the same expression. We refer to the Supplementary Material, Part SM2, for full experimental instructions.

*Results*

Excluding subjects who did not pass the comprehension questions or completed the survey more than once, we remain with N=1,662 subjects. The results, plotted in Figure 5, provide clear evidence that morally nudged subjects give, on average, more than those in the baseline. This is confirmed by linear regression that predicts giving as a function of a dummy variable that represents whether a participant participated in the baseline condition or in the morally nudged condition (Emergency: 17.9% vs 11.7%, p = 0.029; Give for France: 26.1% vs 16.6%, p = 0.001). The coefficients of the regressions reveal that, in the *Emergency* condition, donations among nudged subjects are 39% higher than donations among non-nudged subjects, whereas, in the Give for France condition, donations among nudged subjects are 47% higher than donations among non-nudged subjects. Merging the two conditions (Emergency and Give for France), we find that moral nudges increase charity donations by about 44%, on average.

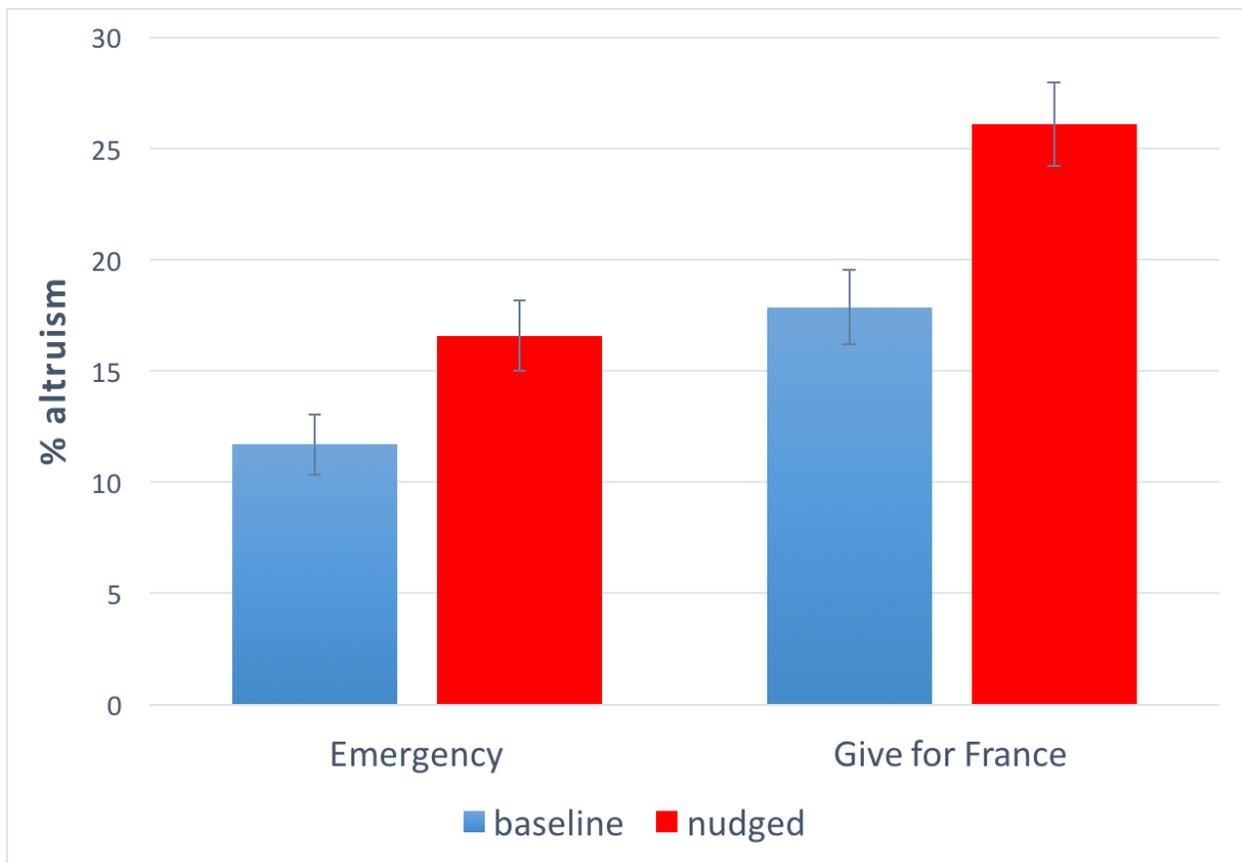

**Figure 5: Moral nudges increase charity donations to real non-profit organisations.** *Results of Study 5 (N=1,662). Morally nudged subjects donate significantly more to humanitarian organisations than those who are not morally nudged. This holds both when the target organisation is Emergency (in which case average donations increase from 5.85c to 8.29c; coeff = 2.28, p = 0.029) and Give for France (in which case average donations increase from 8.93c to 13.04c; coeff = 4.20, p = 0.001). The coefficients are, respectively, 39% and 47% of the baseline donation. Merging the two conditions (Emergency and Give for France) reveals that, overall,*

*nudging subjects increase donations by 44%. The y-axis represents the average percentage of charity donation by subjects in the corresponding condition. Error bars represent +/- SEM.*

**Discussion**

Finding mechanisms to promote pro-social behaviour is fundamental for the well-being of our societies and is more urgent than ever in a time of key global challenges such as resource conservation, climate change, and social inequalities[1-23]. Among these mechanisms, especially important, because easy and cheap to implement, are the so-called *nudges*, that is, mechanisms that can promote pro-social behaviour, without changing people's action space and their economic incentives[24-25].

Here we focused on moral nudges. In our experiments, participants had to decide between a pro-self and a pro-social course of action, but, before making their actual choice, they were asked what *they think is* or *what they think others consider* to be the morally right thing to do. Our data show that these moral nudges have a positive impact on altruistic (Study 1) and cooperative (Study 2) behaviour in economic games. This positive impact is not limited to the choice made immediately after the nudge but persists to a second interaction (Study 3). Moreover, a significant proportion of this effect even spills across contexts: nudging altruism in a first interaction promotes cooperation in a second, non-nudged, interaction, and nudging cooperation in a first interaction promotes altruism in a second, non-nudged, interaction (Study 4). Furthermore, this positive impact is not limited to economic games: nudging altruistic behaviour can be used to successfully increase charity donations (Study 5).

This is an important improvement over previous work. Although it has been repeatedly shown that people tend to follow what they perceive to be the norm in a given context[26-35], there is still a lot of uncertainty about the effects of moral nudges, especially concerning their spillover effects across contexts. This question is particularly important in light of previous literature offering evidence of both positive[46-50] and negative spillovers. In particular, negative spillovers are reported in the literature on moral cleansing, which shows that moral actions are often followed by compensatory immoral actions, suggesting that some moral nudges may backfire[38-42]. Our results show that this is not the case, at least for the moral nudges, the window of time, and the economic decisions considered in this paper. Not only do our moral nudges not backfire, but they even persist over time and spill across contexts.

To the best of our knowledge, only two papers have previously considered moral nudges of the type we have used in this work. Brañas-Garza[74] found that telling DG dictators to "note that he relies on you" increases donations in the laboratory. We go beyond the results of this paper along several dimensions, as we explore the effect of moral nudges, both in the DG and in the PD, we look at its persistence over interactions and across contexts; and we also look at its effect on crowdfunding charity donations to humanitarian organisations. Dal Bó and Dal Bó[33], instead, found that reminding the "Golden Rule" in the middle of a repeated PD with random re-matching in groups of eight people and feedback after each round increases cooperation for a few rounds before eventually vanishing. However, the fact that the interaction is repeated in small groups with feedback after each round implies that the persistence of the increase in cooperation cannot be attributed to the persistence of the nudge with certainty. For example, it could be driven by reputation, if, for example, subjects play a Tit-for-Tat strategy conditional on the strategy they encountered in the previous interaction. We thus go much beyond the results by Dal Bó and Dal Bó[33], in several ways: we look at the effect of moral nudges also in the DG, at its persistence across interactions while giving no feedback about the previous interaction (and therefore ruling out the potential confound of conditional strategies), at its persistence across contexts, and at its effect on crowdfunding charity donations to humanitarian organisations.

Our results have potentially impactful applications for policy design. Several mechanisms to promote pro-social behaviour have been explored in previous work, including giving material reward, such as a t-shirt or a mug, in exchange to a pro-social action[75-76], augmenting donations using matching[77-78], making people's actions observable by others[79-80], informing people about the actions of others to make a social norm salient[81-82], giving gifts *while* asking for a donation, in order to induce a reciprocal feeling of obligation[83-84], soliciting people to cooperate[85-86], and many others (see Kraft-Todd et al[6] for a review). Our results expand this list of mechanisms significantly. We show that moral nudges not only increase pro-social behaviour in the choice immediately after the nudge, but that their effect also persists over time and spills across contexts. Furthermore, compared to these studies, moral nudges have the crucial upside that they are extremely easy and cheap to implement, while still being very effective. For example, Landry et al[87] entered people who donated to a fund-raiser into a lottery to win a personal cash prize, and found a 47% increase in the amount of money raised relative to the control condition with no lottery. Our Study 5 shows that moral nudges produce – at least in the domain of crowdfunding donations in which we used them – essentially the same increase (44%), but are free of cost, that is, they allow to save the money for the prize of the lottery and the time needed to organise it and conduct it. This implies that our moral nudges might be a promising tool for crowdfunding charity donations to humanitarian organisations. This is an especially interesting application, in light of the emergence of crowdfunding websites such as gofundme.com, kickstarter.com, and donationcrowdfunding.com. Of course, the fact that our moral nudges work well in increasing donations for crowdfunding does not imply that they also work well in promoting everyday pro-social behaviour in other parts of life. Exploring the external validity of moral nudges is an important direction for future work.

Our findings also have some limitations. First of all, they show that moral nudges increase pro-social behaviour, persist over time, and spill across contexts, but only within the window of time and choices considered in this paper, that is, two interactions (DG and PD) within the same experiment (i.e., separated by a few minutes). Future work should thus consider interactions separated by a longer time span and a greater number of interactions to explore the boundaries of the effectiveness of moral nudges. Furthermore, it is to be expected that the effectiveness of a moral nudge will not only depend on the particular type of nudge used, but also on the underlying social interactions. Thus, more generally, classifying moral nudges in terms of their effectiveness in different social interactions is an important direction for future research, with many applications to policy design. Second, especially Studies 1-4 use very small stakes, thus it is not clear if and how our results generalise to larger stakes. However, we believe that this is not a major limitation because previous literature found that people's behaviour in AMT experiments using small stakes is essentially equivalent to behaviour in experiments using standard lab stakes both in the DG and in the PD[88-90] (See ref. 90 for a review). Additionally, our first session of Study 5 uses larger stakes and finds results that are in line with the other studies. Here the pro-social choice is relatively expensive, because we asked participants to donate the whole participation fee (50 cents), and we do so at the very end of the survey. This adds to the monetary cost of the altruistic action, a non-monetary utility associated with the sense of deservedness of the participation fee, which decreases the attitude to donate[91]. Even in this "extreme" case, we find a significant effect of the moral nudge (coeff=4.899, p=0.018). Third, our experiments do not allow us to draw conclusions about the psychological underpinnings of our effects. Do people change their behaviour because they are motivated by a sincere desire of doing what they think is the morally right thing to do? Or, alternatively, do they change their behaviour because the experimental setting makes it salient that there is an appropriate thing to do? Indeed, the effect of the moral nudges in our experiments might have been driven by an experimenter demand effect[92]. This is not an issue, however, because susceptibility to experimenter demand effects can be thought of as an instance of norm-sensitivity[93]. Another potential explanation for our results is that they are driven by reputation: it could be that moral nudges make reputation salient, and that this ultimately has the effect of increasing pro-sociality. At this stage, our findings only allow us to conclude that moral nudges can be used effectively to increase pro-social behaviour (cf. especially Study 5). We named them "moral nudges" because they literally refer to morality, but this name is not meant to refer to the underlying psychological mechanism that makes these nudges effective. Understanding the psychological underpinnings of our effect is an interesting direction for future research. Fourth, our experiment does not control for subjects' level of experience in similar AMT experiments. As noted in previous research[70], experienced subjects may decrease effect sizes due to the fact that they are more resistant to experimental manipulations. This implies that the size of the positive effect of moral nudges reported in this work is likely to be a lower bound of its true size. Fifth, Study 5 shows that moral nudges increase charitable donations in the domain of gains, that is, in cases in which participants are asked to donate a sum of money that they have just earned. Future work should explore whether the same nudges work also in the domain of losses, that is, in situations in which participants are nudged to donate a sum of money that they already possess. Sixth, we analysed the effect on pro-sociality of only two moral nudges: a nudge meant to make the personal norm[94] salient ("what do you think is the morally right thing to do?") and a nudge meant to make the descriptive norm[28] salient ("what do you think your society considers to be the morally right thing to do?"). Future research should extend this approach by investigating other ways to activate the same type of norms or even other norms, as for example, prescriptive norms ("what do you think your society thinks that you should do?") or proscriptive norms ("what do you think your society thinks that you should not do?").

Finally, our data also has theoretical applications. We have shown that people's behaviour in one-shot anonymous interactions can be influenced by moral nudges. This is in contrast to the predictions of the standard models proposed by behavioural economists to explain human pro-sociality in one-shot anonymous interactions. According to these models, people have preferences for minimising social inequities[95-96], or maximising social welfare[97], or a combination of both[97]. These preferences are described in terms of the economic outcomes of the available actions and parameters representing the extent to which an individual cares about equity and/or efficiency. Therefore, these models predict that people should be insensitive to cues about the rightness or the wrongness of an action, because these cues do not change the economic outcomes of the available actions. This prediction is not satisfied by our data: cues about what is right in a given situation *can* significantly change people's behaviour. Thus, our results highlight the necessity of going beyond outcome-based preferences, and speak in support of incorporating social norms into people's preferences. This is in line with an emerging strand of research suggesting that people, in their decision-making, strive for balance between maximising their material payoff and doing what they think is the morally right thing to do[34-35,93,98-102].

In sum, we have presented a novel technique to increase pro-social behaviour. Asking people "what's the morally right thing to do?" before they make a choice, makes the morality of an action salient. This promotes pro-social behaviour, and this positive effect persists to a subsequent choice, even when the social context changes. Moreover, it can successfully be used to increase donations to charitable organizations in crowdfunding campaigns. These results hold within the natural domain of our experiments. Future research should explore potential generalisations, with a particular focus to larger stakes, longer time spans and a greater number of interactions.

**Author contributions statement**

**Competing financial interests**